\documentclass[prd,twocolumn,tightenlines,showpacs,nofootinbib,amsfonts,amssymb,amsmath]{revtex4-1}
\usepackage{graphicx}
\begin{document}
\newcommand{\etal}{{\it et al.}}
\newcommand{\bx}{{\bf x}}
\newcommand{\bn}{{\bf n}}
\newcommand{\bk}{{\bf k}}
\newcommand{\dd}{{\rm d}}
\newcommand{\dslash}{D\!\!\!\!/}
\def\ga{\mathrel{\raise.3ex\hbox{$>$\kern-.75em\lower1ex\hbox{$\sim$}}}}
\def\la{\mathrel{\raise.3ex\hbox{$<$\kern-.75em\lower1ex\hbox{$\sim$}}}}
\def\beq{\begin{equation}}
\def\eeq{\end{equation}}

\leftline{UMN--TH--3342/14}

\vskip-2cm
\title{The expected anisotropy in solid inflation}
\author{Nicola Bartolo$^{1}$,  Marco Peloso$^{2}$, 
Angelo Ricciardone$^{1}$, Caner Unal$^{2}$, 
}
\affiliation{
${^1}$ Dipartimento di Fisica e Astronomia �G. Galilei�, \\
Universit\`a degli Studi di Padova, I-35131 Padova (Italy) \\
${^2}$ School of Physics and Astronomy,
University of Minnesota, Minneapolis, 55455 (USA)\\
}
\vspace*{2cm}

\begin{abstract}
Solid inflation is an effective field theory of inflation in which isotropy and homogeneity are accomplished via a specific combination of anisotropic sources (three scalar fields that individually break isotropy).  This results in specific observational signatures  that are not found in standard models of inflation: a non-trivial angular dependence for the squeezed bispectrum, and a possibly long period of anisotropic inflation (to drive inflation, the ``solid'' must be very insensitive to any deformation, and thus background anisotropies are very slowly  erased). In this paper we compute the expected level of statistical anisotropy in the power spectrum of the curvature perturbations of this model. To do so, we account for the classical background values of the three scalar fields that are generated on large (superhorizon) scales during inflation via a random walk sum, as the perturbation modes leave the horizon.
 Such an anisotropy is unavoidably generated, even starting from perfectly isotropic classical initial conditions. 
The expected level of anisotropy is related to the  duration of inflation and to the amplitude of the squeezed bispectrum. 
If this amplitude is close to its current observational limit (so that one of the most interesting predictions of the model can be observed in the near future), we find that  a level of statistical anisotropy $\ga 3\%$ in the power spectrum is to be expected,  if inflation lasted $\ga 20-30$ e-folds more than the final $50-60$ efolds required to generare the CMB modes. We also comment and point out various similarities between solid inflation and models of inflation where a suitable coupling of the inflaton to a vector kinetic term $F^{2}$ gives frozen and scale invariant vector perturbations on superhorizon scales.

\end{abstract}
 \date{July 2014}
 \maketitle

\section{Introduction}

Observations strongly support the inflationary paradigm \cite{Ade:2013uln}, and, in particular, they appear to be in agreement with the simplest single field slow-roll inflationary predictions. However, we are   ignorant on what the correct model of inflation is. A severe  obstacle for inflation is to preserve the flatness of the inflaton potential against radiative corrections. This problem is particularly pressing in models that produce a large gravity wave background, as the flatness needs to be preserved for a large interval of field values \cite{Lyth:1996im}. A well studied  way to address this problem is to impose a shift symmetry broken in a controllable way, see \cite{Pajer:2013fsa} for a recent review. 

The use of symmetries is at the basis of the effective field theory of inflation \cite{Cheung:2007st,Weinberg:2008hq}. As in all effective field theories, one constructs the most general operators of a given order that are compatible with a given symmetry, and one studies their phenomenological implications, with the goal of determining or constraining the coefficients of these operators from the data. In the vast majority of the models, the operators under consideration are invariant under spatial transformations - as suggested by the homogeneity and isotropy of the inflationary background - while the invariance under time translation is slightly broken (compatible with the controllable small breaking of the shift symmetry in the concrete UV realizations that adopt this symmetry). An exception to this is the model of {\it solid inflation}  \cite{Endlich:2012pz}, in which three  scalar fields ${\hat \phi}^i$, that individually break isotropy, are combined to produce a homogeneous and isotropic expansion.\footnote{Another exception being, in a similar way, models of inflation that involve vector fields with a nonvanishing vev.} The scalars are imposed to have orthonormal vevs: 
\begin{equation}
\langle {\hat \phi}^i \rangle = x^i \;\;\;,\;\;\; i=1,2,3 \;. 
\label{solid-bck}
\end{equation}

Such a model was first proposed in \cite{Gruzinov:2004ty} under the name of {\it elastic inflation}, and its properties and its phenomenology have been studied in more details in  \cite{Endlich:2012pz}. \footnote{ In addition to the works that we review below, the perturbations in this model where also studied in ref. \cite{Sitwell:2013kza} and ref. \cite{Akhshik:2014gja}. In particular, ref.  \cite{Akhshik:2014gja} considered the possibility of an anisotropic background,  analogously to  \cite{Bartolo:2013msa}. See also \cite{Ballesteros:2012kv,Chen:2013kta,Balek:2014uua,Balek:2014lza,Pearson:2014iaa} for  recent related works. } 

The reason for the name chosen in  \cite{Endlich:2012pz} is that this model   can also be understood as  a basic effective field theory for a solid. At zeroth order, the properties of a solid are determined by the positions of its fundamental cells. Such positions can be determined by the three scalars ${\hat \phi}^i \left( t, x^i \right)$, whose vevs give  the coordinates at the time $t$ of the cell that at  time $t=0$ was at position $x^i$. The vevs (\ref{solid-bck}) then describe a solid at rest. In the inflationary application of  \cite{Endlich:2012pz} they describe a solid at rest in comoving coordinates. To make the vevs 
(\ref{solid-bck}) compatible with homogeneity and isotropy, ref. \cite{Endlich:2012pz} imposes that the lagrangian of solid inflation is only function of SO(3) invariant combinations of 
\begin{equation}
{\hat B}^{ij} \equiv g^{\mu \nu} \, \partial_\mu {\hat \phi}^i \, \partial_\nu {\hat \phi}^j \;. 
\label{B-ij}
\end{equation} 
Therefore a Friedmann-Lemaitre-Robertson-Walker (FLRW) background is achieved by (i) imposing SO(3) invariance of the theory under rotations  ${\hat \phi}^i \rightarrow R^i_j \, {\hat \phi}_j$, and (ii) tuning the vacuum expectation values (vevs) of the three scalars to be orthonormal.  With these minimal sets of assumptions, solid inflation is a very simple effective scalar field theory of inflation for which FLRW invariance is obtained from a combination of space dependent sources, and for which therefore isotropy is not a fundamental symmetry. Due to its peculiar symmetry breaking, solid inflation has some key phenomenological consequences that differ from those of standard scalar field models of inflation. 

One such consequences is a nontrivial angular dependence in the squeezed limit of the three point functions (the bispectrum)  \cite{Endlich:2012pz,Endlich:2013jia}, which is not present in standard scalar field models. The squeezed bispectrum is characterized by one momentum, say $\vec{k}_3$, much smaller than the other two ($\vec{k}_1 \simeq - \vec{k}_2$) in the correlator. The corresponding mode leaves the horizon before the other two, and it modulates their power spectrum. The directionality of $\vec{k}_3$ imprints an angular dependence to this modulation. However,  in standard scalar field models this angular dependence is proportional to the gradient of the $\vec{k}_3$ mode, and goes to zero in the $\vec{k}_3 \rightarrow 0$ limit $\left(\rm squeezed\; limit \right)$ \cite{Babich:2004gb,Lewis:2011au}. On the contrary,  ${\rm O } \left( \delta {\hat \phi}^3 \right)$ operators in solid inflation have a directionality dependence inherited by (\ref{solid-bck}), which persists also in the squeezed limit. Such dependence has been studied and confronted with the data in \cite{Shiraishi:2013vja,Ade:2013ydc}. 

Another peculiarity of solid inflation is that it can support a prolonged period of anisotropic inflation \cite{Bartolo:2013msa}. It is clear that the medium of solid inflation is a very unconventional solid (as compared to those that we ordinarily interact with), as it is nearly unaffected by the huge inflationary expansion. To sustain inflation, its energy density had to be nearly constant over the inflationary $\geq \left( {\rm e}^{60} \right)^3$ volume increase. This can be enforced by a suitable choice of the lagrangian. Once this is done, stability and gaussianity of the primordial perturbations then require that this medium is very insensitive to any deformation \cite{Endlich:2012pz}  (see eqs.~(\ref{10}) and (\ref{NG-c2})). This is the fundamental reason why, contrary to the generic inflationary expectation  \cite{Wald:1983ky,Maleknejad:2012as},  this medium is unable to quickly erase a background anisotropy \cite{Bartolo:2013msa}. In fact it is possible (and, probably, likely \cite{Bartolo:2013msa}) that the medium of solid inflation is unable to quickly lead to isotropy and homogeneity starting from generic initial conditions. If this is the case, it is possible that solid inflation does not lead to a solution of the isotropy and homogeneity problem. Following the classical background evolution only (as done in 
 \cite{Bartolo:2013msa}) one obtains that isotropy is reached in ${\rm O } \left( \frac{1}{\epsilon} \right)$ e-folds (where $\epsilon \ll 1$ is the standard slow-roll parameter), contrary to the ${\rm O } \left( 1 \right)$ e-folds of the standard models. This leads to the question of whether quantum effects oppose this slow classical isotropization (and, more in general, homogenization), and, possibly, prevail over it.  The present investigation aims to study this issue.   

These  two peculiar properties (namely, the nonstandard squeezed bispectrum and the slow isotropization)
had previously been encountered in models with vector fields during inflation. It is well known that in the standard U(1) invariant case, ${\cal L} = - \frac{1}{4} F^2$, vector fields are not excited by the background expansion, and, if produced by something else, they rapidly decay. To contrast this, a number of models broke the U(1) symmetry in different ways, showing that this can allowed for a long-lived vector vev \cite{Ford:1989me,Turner:1987bw,Ackerman:2007nb,Golovnev:2008cf,Dimopoulos:2008yv}. 
The breaking of the U(1) symmetry introduces an additional polarization that turned out to be a ghost for all these models \cite{hcp}. A more successful alternative is to modify the kinetic term while preserving the U(1) invariance. This can be done for instance by multiplying the standard kinetic term by a function of the inflation, ${\cal L} = - \frac{f \left( \phi \right)}{4} F^2$. A suitable choice of this function can result in (nearly) constant magnetic  \cite{Ratra:1991bn} or electric \cite{Watanabe:2009ct} vev of the vector field. This second choice can support anisotropic inflation.~\footnote{While in the magnetogenesis application  \cite{Ratra:1991bn} the vector field needs to be identified with the standard model photon, this association is not necessary for anisotropic inflation.} The same function also produces frozen scale and (nearly) scale invariant vector perturbations outside the horizon. These perturbations imprint an anisotropy to the power spectrum (PS)  \cite{Dulaney:2010sq,Gumrukcuoglu:2010yc,Watanabe:2010fh} and give rise to a bispectrum that, analogously to that of solid inflation, has a nontrivial angular dependence in the squeezed limit, where it is peaked \cite{Barnaby:2012tk,Bartolo:2012sd}.

In all these studies the vector vev is taken as an arbitrary initial condition, with the exception of \cite{Bartolo:2012sd}. This last work estimated the value of the vev that can be expected in a given inflationary realization. This quantity is related to the infra-red (IR) modes of the vector field. These are the modes that left the horizon between an early moment during inflation (we conventionally denote this moment as the beginning of inflation) and the moment in which the large scales CMB modes left the horizon. To our knowledge, the first work that studied the effect of the vector IR modes is ref. \cite{Demozzi:2009fu}, which showed that if the vector field has a red spectrum, the energy in these IR modes becomes comparable to that of the inflaton, so that their backreaction strongly affects the inflaton dynamics \cite{Kanno:2009ei} (and, typically, suppress the magnetic field production to an unobservable level). It is implicit in \cite{Demozzi:2009fu,Kanno:2009ei} 
 that the IR modes, despite being of wavelength larger than our Hubble patch,  have a physical influence on the background dynamics. 
 
Each IR mode is observed in our patch as a classical homogeneous quantity. Observations can only be affected by the sum of these modes, and this sum is indistinguishable from a spatially constant vev. Consider for instance a very light scalar field in a  de-Sitter (dS) geometry. As well known \cite{Linde:1982uu,Starobinsky:1982ee,Vilenkin:1982wt} (and as we review below), the variance of this field grows with time as $\langle \phi^2 \rangle = \frac{H^3}{4 \pi^2} t = \left( \frac{H}{2 \pi} \right)^2 \, N$, where $H$ is the expansion rate, and $N=H t$ is the number of e-folds. 
Modes that leave the horizon during these $N$ e-folds contribute to this variance, and the $\langle \phi^2 \rangle \propto t$ proportionality is typical of the stochastic addition of these modes (a ``random walk'' sum). 
If $\phi =0$ at the initial time, and if we could observe a large number of realizations of these $N$ e-folds of expansion, and plot the distribution of $\langle \phi \rangle$ obtained in these realizations, we would obtain a gaussian curve with zero mean and  with the above variance. If, as it is the case, we can observe only one realization of these $N$ e-folds, then $\langle \phi \rangle_{\rm 1 realization}$ should be treated as a classical stochastic variable drawn from such gaussian distribution. It is possible that, by mere chance, the IR modes added up to a very typical value in our patch (even a value very close to zero). However, we should expect 
\begin{equation}
\langle  \phi \rangle_{\rm 1 realization,  expected} = \frac{H}{2 \pi} \, N^{1/2} \;, 
\end{equation}
while very different values  are exponentially improbable. 

If anisotropic sources are present, and if the model is constructed in such a way that these sources have a scale invariant spectrum, then their individual vevs break isotropy. Each component of a vector field will in this case be a classical quantity obtained from a gaussian distribution of variance $\propto N$. As a consequence, if this model is realized in nature, the vector during inflation will develop a vev in our Hubble patch that breaks isotropy. Ref. \cite{Bartolo:2012sd} showed that, in the  $- \frac{f \left( \phi \right)}{4} F^2$, the expected vev developed with $\sim 3$ e-folds of inflation is enough to lead to unacceptable violation of statistical isotropy \cite{Kim:2013gka}, if $f \left( \phi \right)$ is chosen so to produce a scale invariant vector spectrum. One could even start with a triplet of vectors, and arrange their vevs to be initially orthonormal
\cite{ArmendarizPicon:2004pm}. This orthonormality will be preserved by the classical equations of motion, if the theory is symmetric under the exchange of these three vectors. There is however no reason why quantum fluctuations should respect this. Each field fluctuates on his own, and the value of $\langle  \vec{A}^{(a)} \rangle_{\rm 1 realization,  expected}$ for each field ($a=1,2,3$) should be understood as originating from three {\it different} drawing from the {\it same statistical  distribution}. In shorts,  the theory is symmetric and isotropic, but any one given realization of the theory in which quantum effects are relevant is not.~\footnote{Isotropy can be preserved if for instance the vectors are massive and their quantum modes are negligible, as in the dark matter model of \cite{Cembranos:2012ng}.} 

A breaking of isotropy from IR effects should be expected also in solid inflation. Even if the theory is SO(3) invariant in field space, and the vevs (\ref{solid-bck}) are imposed at the start, the IR modes will in general add up to values that break the orhonormality of the three vevs.  The goal of the present investigation is to quantify this effect. Specifically,  we point out the similarities with the effects of the IR modes within the 
$f \left( \phi \right) F^2$ inflation models and then we compute the random walk sum of the various individual IR modes. Using the solution of the anisotropic background found in~\cite{Bartolo:2013msa} for the evolution of each IR mode in the super-horizon regime, 
we are able to compute the resulting level of background anisotropy, and the consequent statistical  isotropy breaking in the power spectrum of primordial curvature perturbations. The breaking in the power spectrum is proportional to the background anisotropy, with a proportionality coefficient related to the amplitude of the squeezed bispectrum in 
the model \cite{Endlich:2013jia}. If this bispectrum is closed to its observational limit, so that one of the most characteristic predictions of the model can be observed perhaps already in the next Planck release, we obtain that the IR modes generated in about $20-30$ e-folds of inflation can lead to violation of statistical anisotropy in excess to the current observational limit  \cite{Kim:2013gka}. 

The paper is organized as follows. In Section \ref{sec:solid-review} we review the model of solid inflation \cite{Endlich:2012pz}, its bispectrum  \cite{Endlich:2012pz,Endlich:2013jia}, and its Bianchi-I solution   \cite{Bartolo:2013msa}. In Section \ref{sec:IR} we compute the IR perturbations of the three fields ${\hat \phi}^i$.  In Sec.~\ref{sec:smallV} we show that the contribution to the IR modes from vector perturbations is subdominant w.r.t to the scalar (longitudinal) ones. In Section \ref{sec:aniso}  and in Sec.~\ref{CMB} we quantify the corresponding breaking of statistical isotropy studying its dependence on the parameters of the model. In Sec.~\ref{conclusion} we comment and interpret our final results.

\vspace{1cm}

\section{The model of solid inflation and two background solutions}
\label{sec:solid-review}

As mentioned in the Introduction, the lagrangian of solid inflation  is function of  SO(3) invariant combinations of the ${\hat B}_{ij}$ elements defined in (\ref{B-ij}). This guarantees that the vevs (\ref{solid-bck}) are compatible with the FLRW background 
\begin{equation}
d s^2 = - d t^2 + a^2 \left( t \right) \, d \vec{x}^2 \;.  
\label{FRW}
\end{equation}
Only three such independent combinations exist, that in   \cite{Endlich:2012pz} are chosen to be 
\begin{equation}
{\hat X} \equiv {\rm tr } \, {\hat B} = {\hat B}_{ii} \;,\; 
{\hat Y} \equiv \frac{{\rm tr } \, {\hat B}^2}{\left( {\rm tr } \, {\hat B } \right)^2} \;,\; 
{\hat Z} \equiv \frac{{\rm tr } \, {\hat B}^3}{\left( {\rm tr } \, {\hat B} \right)^3} \,. 
\end{equation} 
Denoting the action of solid inflation as 
\begin{equation}
S = \int d^4 x \sqrt{-g} \left\{ \frac{M_p^2}{2} R + F \left[ {\hat X} ,\, {\hat Y} ,\, {\hat Z} \right] \right\} \;, 
\label{action}
\end{equation} 
 one obtains, on the background configuration 
(\ref{solid-bck}) and (\ref{FRW}), the energy density and pressure
\begin{equation}
\rho = - F \;,\; p = F - \frac{2}{a^2} F_X \,, 
\end{equation}
where the suffix on $F$ denotes functional derivative (we see that both $F,F_X < 0$ are required). To have inflation, one requires
\begin{equation}
\epsilon \equiv - \frac{\dot{H}}{H^2} = \frac{X \, F_X}{F} \ll 1  \;. 
\label{epsilon}
\end{equation} 

We note that $X,Y,Z$ are normalized in such a way that only $X$ depends on the volume factor of the universe. The slow roll condition (\ref{epsilon}) can be understood as the requirement that the solid is extremely intensity of volume deformations. The fact that $Y$ and $Z$ are insensitive to the volume  immediately explains why derivatives of $F$ with respect to $Y$ and $Z$ do not enter in the expression for the pressure, and are unconstrained by the slow roll requirement. Nonetheless also these derivatives are constrained. At zeroth order in slow roll, and in the deep sub-horizon regime, the sound speed of the scalar (L=longitudinal) and vector (T=transverse)
 perturbations of the solid are given by 
\begin{equation}
c_L^2 \simeq \frac{1}{3} + \frac{8}{9} \, \frac{F_Y + F_Z}{X \, F_X} \;,\; c_T^2 \simeq \frac{3}{4} \left( 1 + c_L^2 \right) \;, 
\label{cL-cT}
 \end{equation}
so that the requirements $0 < c^2 < 1$ translate into 
\begin{equation}
\label{10}
0 < F_Y + F_Z < \frac{3}{8} \, \epsilon \vert F \vert \;. 
\end{equation}
and therefore $F_Y + F_Z \ll \vert F \vert$. 

A further study of the perturbations in this model shows that the individual $F_Y$ and $F_Z$ derivatives also need to be small. This can be seen from the study of non-gaussianity (NG) in the model. As we remarked in the Introduction, the bispectrum $B_\zeta$ of solid inflation is enhanced in the squeezed limit, where it shows a nontrivial angular dependence. It is convenient to use the parametrization for the bispectrum introduced in \cite{Shiraishi:2013vja}: 
\begin{equation}
B_\zeta \left( \vec{k}_1 , \vec{k}_2 , \vec{k}_3 \right) = \sum_L c_L \, P_L \left( \mu_{12} \right) P_\zeta \left( k_1 \right) P_\zeta \left( k_2 \right) + \; 2 \; {\rm perm.} \;, 
\label{skpb}
\end{equation}
where $\vec{k}_i$ are the comoving momenta of the modes in the correlator, $P_L$ are the Legendre polynomials, $P_\zeta$ is the PS, and finally $\mu_{\rm ij}$ is the cosine of the angle between $\vec{k}_i$ and $\vec{k}_j$.~\footnote{We stress that standard models of scalar field inflation have $c_0 \neq 0$ only \cite{Lewis:2011au} in the squeezed limit. Higher order terms are a signature of sources that break isotropy \cite{Shiraishi:2013vja}, as was first obtained in presence of vector fields \cite{Barnaby:2012tk,Bartolo:2012sd}, and later in solid inflation  \cite{Endlich:2012pz}.} Solid inflation gives  \cite{Endlich:2012pz,Endlich:2013jia} 
\begin{equation}
c_2 = \frac{20 F_Y}{9 F \epsilon c_L^2} = 3.8 \pm 27.8 \; \left( 68\% \, {\rm CL} \right) \;, 
\label{NG-c2}
\end{equation} 
where the numerical value is the Planck limit \cite{Shiraishi:2013vja,Ade:2013ydc}. We therefore see that all $X \, \vert F_X \vert, F_Y, F_Z$ need to be $\ll \vert F \vert$. 

Therefore the medium of solid inflation needs to respond very weakly to any geometry deformation. Starting from this observation, ref. \cite{Bartolo:2013msa} showed that the model allows for a prolonged anisotropic inflationary expansion, at odds with standard inflationary models, in which any initial anisotropy is erased within a few Hubble times \cite{Wald:1983ky,Maleknejad:2012as}. For simplicity, ref. \cite{Bartolo:2013msa} considered a Bianchi-I universe with planar symmetry in the $y-z$ directions: 
\begin{equation}
d s^2 = - d t^2 + {\rm e}^{2 \alpha \left( t \right) - 4 \sigma \left( t \right)} d x^2 +  {\rm e}^{2 \alpha \left( t \right) + 2 \sigma \left( t \right)} \left[ d y^2 + d z^2 \right] \,, 
\label{bianchi-I} 
\end{equation} 
and showed that the anisotropy evolves as 
\begin{equation}
\sigma \left( t \right) = \sigma_1 \, {\rm e}^{- 3 \int^t H d t' } + \sigma_2 \, {\rm e}^{- \int^t \left( 1 + c_L^2 \right) \,  \epsilon H d t'} \,, 
\label{sigma-class}
\end{equation} 
where $\sigma_{1,2}$ are integration constants. While the first mode exhibits the typical fast decrease of the anisotropy, the decrease of the second mode is suppressed by the slow roll parameter $\epsilon \ll 1$, so that any initial anisotropy can survive for ${\rm O } \left( \frac{1}{\epsilon} \right)$ e-folds during inflation. 

While ref.  \cite{Bartolo:2013msa}  studied the background evolution starting from an anisotropic initial condition ($\sigma_{1,2} \neq 0$), in the next sections we study how the anisotropy is developed by the generation of long wavelength modes during inflation.

\section{IR sum of the perturbations of the scalar fields in solid inflation}
\label{sec:IR}

In the  $f(\phi)F^{2}$ mechanism, the kinetic term of the vector field is multiplied by  an ad-hoc function of the inflaton field $\phi$, appropriately chosen so to allow for a constant  electric or magnetic component of the vector field (see \cite{Bartolo:2012sd} for a general discussion). This also result in scale invariant and constant vector perturbations $\delta \vec{A}$ at super-horizon scales. Even in absence of any other coupling, the same $f(\phi)F^{2}$ operator responsible for the constant background $\vec{A}^0$ also gives rise to a coupling  $\propto\vec{A}^0 \cdot \delta \vec{A} \, \delta \phi$, which imprints statistical anisotropy in the inflation perturbations  \cite{Dulaney:2010sq,Gumrukcuoglu:2010yc,Watanabe:2010fh}. 

As shown in  \cite{Bartolo:2012sd}, even if one choses initial conditions such that $\vec{A}^0 =0$ at some given moment during inflation, the long wavelength modes of the vector fields generated during inflation add up (in a stochastic way) to a coherent classical background, that is completely indistinguishable from a homogeneous field  $\vec{A}^0$ by modes on much smaller scales. Therefore the vector field unavoidably generates background anisotropy, and statistical anisotropy of the primordial perturbations.  We now show that the role played by the vector vev in the $f(\phi)F^{2}$ mechanism is here played by a difference in the background gradients of the three fields. 

In the original solid inflation paper \cite{Endlich:2012pz}, the classical background configuration (\ref{solid-bck}) is assumed. We assume that the scalar fields profile agrees with (\ref{solid-bck}) at some early time $t_{\rm in}$ during inflation, that we conventionally denote as the start of inflation. We denote by $N_{\rm tot}$ the number of e-folds of inflationary expansion from this moment to the end of inflation. In general, this quantity is greater than (and, possibly, much greater than) the number of e-folds before the end of inflation at which the large scale CMB modes left the horizon. We denote this latter quantity as $N_{\rm CMB}$, with the standard expectation that $N_{\rm CMB} \simeq 50-60$.

We denote as IR modes the modes of  $\delta {\hat \phi}^i$ that leave the horizon in the ``initial'' $N_{\rm tot} - N_{\rm CMB}$ e-folds of inflation. These modes are seen as a classical homogeneous  background by modes of CMB and smaller scales, and, for all purposes, are indistinguishable from a homogeneous background in our Hubble patch. These modes  have already become classical by the time the CMB modes are produced, and should be added up with (\ref{solid-bck}) to provide the background on which the CMB modes are produced.~\footnote{With this understanding, there is no physical reason why super-horizon modes should be absent at any given moment, as we have assumed to be the case by imposing that the background configuration is precisely given by (\ref{solid-bck}) at $t=t_{\rm in}$. If present, these modes will in general increase the   anisotropy with respect to the amount that we compute here. Therefore, our result is in general a lower bound estimate for the amount of anisotropy that should be expected after $N_{\rm tot}-N_{\rm CMB}$ e-folds of inflation.}

To compute this IR contribution, we decompose the perturbations as in  \cite{Endlich:2012pz}: 
\begin{equation}
\delta {\hat \phi}^i = \frac{i k^i}{k} {\hat \pi}_{L,\vec{k}}(t)+ {\hat \pi}_{T,\vec{k}}^i (t) \;,
\label{solid-perts}
\end{equation}
and we adopt the 
\begin{equation}
{\hat f}_{\vec{k}} \left( t  \right) = \int \frac{d^3 x}{\left( 2 \pi \right)^{3/2}} \, {\rm e}^{-i \vec{x} \cdot \vec{k}} \, {\hat f} \left( t , \vec{x} \right) \;, 
\label{FT}
\end{equation}
convention for the Fourier transform. 

The mode ${\hat \pi}_L$ in (\ref{solid-perts}) is a scalar perturbation. Scalar perturbations are customarily characterized by the gauge invariant variable ${\hat \zeta}$, that represents the curvature perturbations on uniform density hypersurfaces. In spatially flat gauge ($\delta g_{ij,{\rm scalar}} = 0$), the longitudinal mode ${\hat \pi}_L$ is related to ${\hat \zeta}$ by  \cite{Endlich:2012pz}: 
\begin{equation}
{\hat \zeta} \equiv-H \, \frac{\delta {\hat \rho}}{\dot{\rho}} = -\frac{k}{3} \, {\hat \pi}_{L} \;. 
\label{zeta}
\end{equation}

The modes ${\hat \pi}_i^T$ in (\ref{solid-perts}) are vector perturbations, and they are transverse. We decompose them in two polarizations:
\begin{equation}
{\hat \pi}_{T,\vec{k}}^i  = \sum_{\lambda=1}^2 \epsilon^i_{\lambda,{\hat k}} \left( {\hat k} \right) \, \frac{{\hat T}_{\lambda,\vec{k}} }{k} \;,  
\label{vector-deco}
\end{equation}
where $k_i \, \epsilon^i_\lambda \left( {\hat k} \right) =0$ and $ \epsilon^i_\lambda \left( {\hat k} \right) 
 \epsilon^i_{\lambda'} \left( {\hat k} \right) = \delta_{\lambda \lambda'} $. The $\frac{1}{k}$ factor has been inserted in 
(\ref{vector-deco}) since the modes $T_\lambda$ are those with a (nearly) scale invariant power spectrum. 

%

We define the PS as usual: 
\begin{eqnarray}
\left\langle {\hat \zeta}_{\vec{k}} \,  {\hat \zeta}_{\vec{p}} \right\rangle & \equiv & \frac{2 \pi^2}{k^3} \, \delta^{(3)} \left( \vec{k} + \vec{p} \right) \, P_\zeta \left( k \right) \;, \nonumber\\ 
\left\langle {\hat T}_{\lambda,\vec{k}} \,  {\hat T}_{\sigma,\vec{p}} \right\rangle & \equiv & \frac{2 \pi^2}{k^3} \, \delta^{(3)} \left( \vec{k} + \vec{p} \right) \, \delta_{\lambda \sigma} \, P_\lambda \left( k \right) \;, 
\label{PS}
\end{eqnarray} 
Under the assumption that $\zeta$ coincides with the observed cosmological perturbations, we write 
\begin{equation}
P_\zeta = P_* \, \left( \frac{k}{k_*} \right)^{n_s-1} \;, 
\label{PS-param}
\end{equation}
where $k_*=0.05 \, {\rm Mpc}^{-1}$ is the Planck pivot scale, while we use the central values $P_* = 2.45 \cdot 10^{-9}$ and $n_s = 0.96$ \cite{Ade:2013uln} in the our estimates below. The PS of the vector modes is related to the scalar one by  \cite{Endlich:2012pz} 
\begin{equation}
P_\lambda = \frac{9 c_L^5}{c_T^5} \, P_\zeta \;, 
\label{PT-PL}
\end{equation}
where we recall that $c_L$ and $c_T$ are the sound speed of the scalar and vector modes, respectively, see eqs. (\ref{cL-cT}). 

The perturbations (\ref{solid-perts}) modify the field profiles (\ref{solid-bck}) into 
\begin{eqnarray}
{\hat \phi^i } &=& 
  x^i  \!\! + \!\! \int_{k_{\rm in}}^{k_{\rm CMB}} \!\!\!\! \frac{d^{3}k}{(2\pi)^{3/2}} e^{i \vec{k} \cdot \vec{x} }  \left( - 3  \cfrac{i k^{i}}{k^{2}} \hat{\zeta}_{\vec{k}} \left(t\right)+  \epsilon^i_\lambda \left( {\hat k} \right) \, \frac{{\hat T}_{\lambda,\vec{k}} \left( t  \right)}{k}  \right)    \nonumber\\
&& + \int_{k_{\rm CMB}} \!\! \frac{d^{3}k}{(2\pi)^{3/2}} e^{i \vec{k} \cdot \vec{x} }  \left( - 3  \cfrac{i k^{i}}{k^{2}} \hat{\zeta}_{\vec{k}} \left(t\right)+  \epsilon^i_\lambda \left( {\hat k} \right) \, \frac{{\hat T}_{\lambda,\vec{k}} \left( t  \right)}{k}  \right)  \,, \nonumber\\ 
\label{phi-tot}
\end{eqnarray}
where $k_{\rm in}$ is the comoving momentum of the modes that left the horizon at $t_{\rm in}$ (in our convention, they are the modes  leaving the horizon $N_{\rm tot}$ e-folds before the end of inflation), while $k_{\rm CMB}$ is the comoving momentum of the large scales CMB modes (those leaving the horizon $N_{\rm CMB}$ e-folds before the end of inflation). 

As we already discussed, for what concerns observations within our Hubble patch, the classical inflationary scalar field background is not given by (\ref{solid-bck}) but rather by the vev of 
\begin{eqnarray}
&& \!\!\!\! 
{\hat \phi}^i_{\rm classical} = x^i + {\hat \phi}^i_{\rm IR} \,, \nonumber\\ 
&& \!\!\!\! 
 {\hat \phi}^i_{\rm IR} =  - 3 \int_{k_{\rm in}}^{k_{\rm CMB}} \!\!\!\! \frac{d^{3}k}{(2\pi)^{3/2}} e^{i \vec{k} \cdot \vec{x} } \; 
\left(  \cfrac{i k^{i}}{k^{2}} \hat{\zeta}_{\vec{k}} \left(t\right) 
- \epsilon^i_\lambda \left( {\hat k} \right) \, \frac{{\hat T}_{\lambda,\vec{k}} \left( t  \right)}{3 \, k}  \right)  \,, \nonumber\\ 
\label{classical} 
\end{eqnarray} 
where $ {\hat \phi}^i_{\rm IR} $ is the sum of IR modes. By the time the CMB modes are produced, these IR modes have left the horizon and have become classical. If we could observe many realization of the first $N_{\rm tot} - N_{\rm CMB}$ e-folds of inflation, and compute the distribution of the IR modes in all these realizations, we would observe a (nearly) gaussian distribution with the variance related to the PS (\ref{PS}). However, we observe only one realization of the first $N_{\rm tot} - N_{\rm CMB}$ e-folds of inflation, and therefore  $ {\hat \phi}^i_{\rm IR} $ needs to be treated as a stochastic classical quantity, drawn by this (nearly) gaussian distribution. 

In general, the IR contribution in  (\ref{classical}) leads to statistical anisotropy. For this reason, we need to assume that it is small (the goal of this study is to quantify this contribution, and to study under which conditions it is indeed sufficiently small). Under this assumption, we can Taylor expand the sum (\ref{classical}) as:  
\begin{equation}
{\hat \phi}_{\rm classical}^i = x^i + \left( {\hat \Theta}_0 \right)^i + \left(  {\hat \Theta}_1 \right)^i_j \, x^j + \left( {\hat \Theta}_2 \right)^i_{jl} \, x^j \, x^l + \dots \,, 
\label{taylor}  
\end{equation} 
where $\vec{x} = 0$ is a given point in the universe (say, our current location).

We recall that the scalar fields enter in the lagrangian of solid inflation in the form  ${\hat B}^{ij} \equiv g^{\mu\nu} \, \partial_{\mu} \, {\hat \phi}^i \, \partial_{\nu} \, {\hat \phi}^j$. So, the quantity ${\hat \Theta}_0$ in (\ref{taylor}) is unphysical and we disregard it. The linear and quadratic terms are instead physical, and they are given by 
\begin{eqnarray}
&& \left( {\hat \Theta_1 } \right)^i_j = \frac{\partial \, {\hat \phi}_{\rm IR}^i }{\partial x^j } \, \Big\vert_{\vec{x} = 0} 
\nonumber\\ 
&& \;\;\;\; =   3 \int_{k_{\rm in}}^{k_{\rm CMB}} \!\!\!\! \frac{d^{3}k}{(2\pi)^{3/2}}  \; 
\left(  \frac{k^i k_j}{k^{2}} \hat{\zeta}_{\vec{k}} \left(t\right) 
+i \epsilon^i_\lambda \left( {\hat k} \right) k_j \, \frac{{\hat T}_{\lambda,\vec{k}} \left( t  \right)}{3 \, k}  \right)  \;, \nonumber\\ 
&& \left( {\hat \Theta_2 } \right)^i_{jl} = \frac{\partial^2 \, {\hat \phi}_{\rm IR}^i }{\partial x^j \partial x^l } \, \Big\vert_{\vec{x} = 0} 
\nonumber\\ 
&& \;\;\;\; =   3 \int_{k_{\rm in}}^{k_{\rm CMB}} \!\!\!\! \frac{d^{3}k}{(2\pi)^{3/2}}  \; 
\left(  \frac{i k^i k_j k_l}{k^{2}} \hat{\zeta}_{\vec{k}} \left(t\right) 
- \epsilon^i_\lambda \left( {\hat k} \right) k_j k_l \, \frac{{\hat T}_{\lambda,\vec{k}} \left( t  \right)}{3 \, k}  \right)  \;.  \nonumber\\ 
\label{theta-12}
\end{eqnarray}

It is instructive to compare this to the IR contribution to the classical profile from the IR modes in the case of standard scalar field inflation.~\footnote{An analogous discussion applies for the profile of a generic field with a scale invariant spectrum during inflation.} Recall that in this case,  in spatially flat gauge, ${\hat \zeta} = - \frac{H}{\dot{\phi}^0} \delta {\hat \phi}$, where $\phi^0$ is the zero mode. Keeping into account the IR contribution to the classical profile gives 
\begin{equation}
{\hat \phi}_{\rm classical}^{\rm standard} = \phi^0 \left( t \right) - \frac{\dot{\phi^0}}{H} 
 \int_{k_{\rm in}}^{k_{\rm CMB}}   \frac{d^3 k}{\left( 2 \pi \right)^{3/2} } \, {\rm e}^{i \vec{k} \cdot \vec{x}} \, {\hat \zeta}_{\vec{k}} \left( t  \right) \;, 
\end{equation}
which we also Taylor expand 
\begin{equation}
{\hat  \phi}_{\rm classical}^{\rm standard} = \phi^0 + \left( {\hat \Theta}_0^{\rm standard} \right) + \left(  {\hat \Theta}_1^{\rm standard} \right)_j \, x^j + \dots \,. 
\label{taylor-standard}
\end{equation} 
resulting in 
\begin{eqnarray}
{\hat \Theta}_0^{\rm standard}  &=& 
 - \frac{\dot{\phi^0}}{H} \, \int_{k_{\rm in}}^{k_{\rm CMB}}  
 \frac{d^3 k}{\left( 2 \pi \right)^{3/2} }  \, {\hat \zeta}_{\vec{k}} \left( t  \right) \;, 
\nonumber\\ 
\left( {\hat \Theta}_1^{\rm standard} \right)_j &=& 
- i \frac{\dot{\phi^0}}{H}  \, \int_{k_{\rm in}}^{k_{\rm CMB}}  
 \frac{d^3 k}{\left( 2 \pi \right)^{3/2} } \, k_j \, {\hat \zeta}_{\vec{k}} \left( t  \right)\,. 
\label{theta-12-stnd}
\end{eqnarray} 

As discussed above, the operators ${\hat \Theta}_i$ are classical stochastic variables drawn by (nearly) gaussian distributions with zero mean and variance given by 
\begin{eqnarray}
\langle {\hat \Theta}_0^{\rm standard} \,  {\hat \Theta}_0^{\rm standard} \rangle &=& \left( \frac{\dot{\phi^0}}{H} \right)^2 \, 
 \int_{k_{\rm in}}^{k_{\rm CMB}}  \frac{d k}{k} \, P_\zeta \left( k \right) \;, \nonumber\\ 
\langle \left( {\hat \Theta}_1^{\rm standard} \right)_i \,  \left( {\hat \Theta}_1^{\rm standard} \right)_j  \rangle &=& \frac{\delta_{ij}}{3} \, \left( \frac{\dot{\phi^0}}{H} \right)^2  
\!\!\! \int_{k_{\rm in}}^{k_{\rm CMB}}  \!\!\! d k \, k \, P_\zeta \left( k \right) \;. \nonumber\\ 
\label{T0-T1-stnd-integral}
\end{eqnarray} 

In this discussion we disregard the small departure from scale invariance, namely we set $n_s=1$ in (\ref{PS-param}) (the effect of $n_s \neq 1$ will be properly taken into account in the following sections). Then, from the standard result 
\begin{equation}
P_\zeta^{\rm standard} = \left( \frac{H^2}{2 \pi \dot{\phi}^0 } \right)^2 \;\;\;\;,\;\;\;\; \left( n_s = 1 \right) \;, 
\end{equation} 
we obtain 
\begin{eqnarray}
\langle {\hat \Theta}_0^{\rm standard} \,  {\hat \Theta}_0^{\rm standard} \rangle &=& \left( \frac{H}{2 \pi} \right)^2 \left( N_{\rm tot} - N_{\rm CMB} \right) \,, \nonumber\\ 
\langle \left( {\hat \Theta}_1^{\rm standard} \right)_i \,  \left( {\hat \Theta}_1^{\rm standard} \right)_j  \rangle &=& 
\frac{\delta_{ij}}{6} \left( \frac{H}{2 \pi} \right)^2 \, k_{\rm CMB}^2 \,. 
\label{T0-T1-stnd}
\end{eqnarray} 

Once inserted in (\ref{taylor-standard}), the first line in (\ref{T0-T1-stnd}) reproduces the well known result 
 \cite{Linde:1982uu,Starobinsky:1982ee,Vilenkin:1982wt} that the variance of the scalar field grows linearly during inflation (as the different IR modes add randomly to each other, their sum grows in the typical ``random walk'' manner) 
\begin{equation} 
\left\langle \left( {\hat \phi}_{\rm classical}^{\rm standard} \right)^2 \right\rangle = \frac{H^3}{4 \pi^2} \, t \;. 
\label{variance-stnd}
\end{equation}
The second line in (\ref{T0-T1-stnd}) shows instead that the scalar field gradient is typically negligible, as it leads to a difference $\Delta {\hat \phi}_{\rm classical} = {\rm O } \left( H \right) $ in our observable patch. This is negligible if $H \ll \phi^0$, which we assume to be the case. 

Despite relevant for the background evolution during inflation (particularly in the self-reproducing regime, in which the quantum diffusion (\ref{variance-stnd}) can even dominate over the classical evolution \cite{Linde:2005ht}) for a single homogeneous scalar field the quantity ${\hat \Theta}_0^{\rm standard}$ cannot be observationally distinguished from the zero mode value $\phi^0$, and indeed $\phi^0$ can simply be rescaled to include the IR contribution in our observable patch. 

Apart from the numerical coefficients, the different behavior with time of the two operators  in (\ref{T0-T1-stnd}) simply follows from the momentum-dependence of the integrands in (\ref{T0-T1-stnd-integral}). 
By comparing (\ref{theta-12}) with (\ref{theta-12-stnd}), we see that ${\hat \Theta}_1$ in solid inflation scales as ${\hat \Theta}_0^{\rm standard}$, while  ${\hat \Theta}_2$ in solid inflation scales as ${\hat \Theta}_1^{\rm standard}$. We therefore see that the quadratic terms in (\ref{taylor}) are negligible, while the linear terms can potentially be relevant. Contrary to the standard case of a single homogeneous scalar field, these terms  lead to violation of statistical isotropy (compare the situation of a homogeneous scalar field and a homogeneous vector field in a given region. The homogeneous scalar field is simply a number, while a homogeneous vector  points in a given direction and breaks isotropy; consider for instance the case of a constant electric field in a parallel-plate capacitor).  We quantify this effect in the next section.

\section{Small vector perturbations contribution to the IR sum} 
\label{sec:smallV}

As discussed in the previous section, we consider the Taylor expansion (\ref{taylor}), with ${\hat \Theta}_0 = 0$ (as this quantity is unphysical) and  ${\hat \Theta}_2 = 0$ (as this quantity is small). From the results (\ref{PS}), (\ref{PT-PL}), and (\ref{theta-12}), we obtain 
\begin{eqnarray} 
&& \left\langle \left( {\hat \Theta}_1 \right)^i_j \, \left( {\hat \Theta}_1 \right)^m_n  \right\rangle 
= \frac{3}{5} \Bigg[ \delta_{ij} \delta_{mn} +  \delta_{im} \delta_{jn} +  \delta_{in} \delta_{jm} \nonumber\\ 
&& \quad\quad + \frac{c_L^5}{c_T^5} \left( 4  \delta_{ij} \delta_{mn} -  \delta_{im} \delta_{jn} -  \delta_{in} \delta_{jm} \right) \Bigg]  \int_{k_{\rm in}}^{k_{\rm CMB}}  \!\!\! \frac{d k}{k}  \, P_\zeta \left( k \right) \;. \nonumber\\ 
\end{eqnarray}

The terms in this expression $\propto  \frac{c_L^5}{c_T^5} $ originate from the vector modes. From the relation between the two sound speeds (namely, the second  eq. in (\ref{cL-cT})), we find that 
$ \frac{c_L^5}{c_T^5}  \leq 6.4\% $ (the maximum is reached at $c_T = 1$). For this reason, in our estimates we disregard the contribution from the vector modes to the IR sum. 

With only the scalar contribution included, $\left( {\hat  \Theta_1} \right)^i_j$ is symmetric in $i \leftrightarrow j$. As a consequence ${\hat  \Theta_1}$ can be diagonalized with a simultaneous rotation of the fields and of the spatial coordinates. Therefore, as long as the vector modes and higher order terms in the Taylor expansion are disregarded, we can write, without loss of generality
\begin{eqnarray}
&& {\hat \phi}^1 = \left( 1 + {\hat \theta}_1 \right) x \;,\; 
{\hat \phi}^2 = \left( 1 + {\hat \theta}_2 \right) y \;,\; 
{\hat \phi}^3 = \left( 1 + {\hat \theta}_3 \right) z \;,\; \nonumber\\ 
&& {\hat \theta}_i =  3 \int_{k_{\rm in}}^{k_{\rm CMB}} \!\!\!\! \frac{d^{3}k}{(2\pi)^{3/2}}  \; 
{\hat k}_i^2 \;  \hat{\zeta}_{\vec{k}} \left(t\right) \;, 
\label{phi-noV}
\end{eqnarray} 
which are the expressions that we use in the following computations. 

\section{Anisotropy from the IR modes}
\label{sec:aniso}

We assume that the scalar field vevs are given by (\ref{solid-bck}) and that the background is isotropic at the ``initial'' time of inflation $t_{\rm in}$. Every time interval $d t$, some modes of $\delta {\hat \phi}^i$ leave the horizon and become classical. This modes will generate a background anisotropy, that we compute in this Section. This anisotropy will then evolve classically according to the classical evolution studied in  \cite{Bartolo:2013msa}. As this evolution proceeds, other modes will exit the horizon, and give contribution to the anisotropy. Therefore, each mode gives a contribution to the anisotropy that starts when this mode leaves the horizon (with a stochastic amplitude that must follow from (\ref{phi-noV}), and which then evolves classically). In the linear regime (which is appropriate, since we know that the final anisotropy needs to be small) these contributions simply add up independently, so we can focus on the contribution of the anisotropy from each individual mode, and then add up. 

Let us therefore focus on a single mode with wavenumber $\vec{k}$, and on the corresponding scalar perturbations ${\hat \zeta }_{\vec{k}}$. We first compute the relation between ${\hat \zeta}_{\vec{k}}$ at horizon crossing and its contribution to the background anisotropy $d {\hat \sigma}_{\vec{k}}$ at that moment. We recall that we are disregarding the subdominant contribution from vector perturbations (see the previous section). We can first of all see that metric perturbations have a negligible effect in this computation. We are working in the spatially flat gauge, $\delta g_{ij} = 0$. The scalar mode ${\hat \zeta}$ not only perturbs $\delta {\hat \phi}^i$, but also $\delta g_{00}$ and $\delta g_{0i}$. However, these latter contributions are slow-roll suppressed, as can be seen from the computation shown in  \cite{Bartolo:2013msa} (see for instance eq. (18) of that work). As a consequence, their contribution to the anisotropy is negligible from the contribution that originates from $\delta {\hat \phi}^i$, and that we now compute.

%
%

To simplify this computation, from now on we assume a SO(2) symmetry between the $y$ and $z$ direction, and set ${\hat \phi}^2 = {\hat \phi}^3$. This assumption is not correct, since, as we remarked, the quantum fluctuations of the the different fields do not respect any symmetry. We do so only for computational simplicity, and to be able to use the results of  \cite{Bartolo:2013msa}, where such a symmetry was also imposed. This will result in a lower bound estimate for the overall anisotropy (since we are artificially setting to zero some of its components). We believe that this estimate will still be very accurate. In practice, the three vevs $\langle \delta {\hat \phi}_{\rm IR}^1 \rangle, \; \langle \delta {\hat \phi}_{\rm IR}^2 \rangle , \langle \delta {\hat \phi}_{\rm IR}^3 \rangle$ generated by the IR sums will be hierarchical, and we can assume that  $\langle \delta {\hat \phi}_{\rm IR}^1 \rangle$ is the largest of them, without loss of generality. The background anisotropy will be mostly determined by that field, and the $x-$ direction will be the most anisotropic. We could even set to zero the sum of the IR fields in the other two directions, without changing the leading order phenomenological signature of the anisotropy. Rather than setting $ \langle \delta {\hat \phi}_{\rm IR}^2 \rangle = \langle \delta {\hat \phi}_{\rm IR}^3 \rangle = 0$, we set them to a common generic value, as this can be easily included in the computation. 

We can now compute the contributions to the anisotropy from the mode ${\hat \zeta}_{\vec{k}}$, as this modes leaves the horizon at the time $t_k$. Just before the modes leave the horizon, the scalar fields and metric are
\begin{eqnarray}
&& {\hat \phi}^1 = \left( 1 + {\hat \theta}_1 \right) x \;, \nonumber\\ 
&& {\hat \phi}^2 = \left( 1 + {\hat \theta}_2 \right) y \;,\; 
{\hat \phi}^3 = \left( 1 + {\hat \theta}_2 \right) z \;,\; \nonumber\\ 
&&  ds^2 =  - d t^2 + {\rm e}^{2 \alpha  - 4 \sigma } d x^2 +  {\rm e}^{2 \alpha  + 2 \sigma } \left[ d y^2 + d z^2 \right] \,, \nonumber\\ 
\label{before}
\end{eqnarray} 
where the SO(2) symmetry in the $y-z$ plane has been imposed, and where the anisotropy is due to all the IR modes that have left the horizon before  ${\hat \zeta}_{\vec{k}}$. 

Just after the mode ${\hat \zeta}_{\vec{k}}$ leaves the horizon, we have instead 
\begin{eqnarray}
&& {\hat \phi}^1 = \left( 1 + {\hat \theta}_1 + d {\hat \theta}_1 \right) x \;, \nonumber\\ 
&& {\hat \phi}^2 = \left( 1 + {\hat \theta}_2  + d {\hat \theta}_2 \right) y \;,\; 
{\hat \phi}^3 = \left( 1 + {\hat \theta}_2  + d {\hat \theta}_2 \right) z \;,\; \nonumber\\ 
&&  ds^2 =  - d t^2 + {\rm e}^{2 \alpha  - 4 \sigma } d x^2 +  {\rm e}^{2 \alpha  + 2 \sigma } \left[ d y^2 + d z^2 \right] \,, \nonumber\\ 
\label{after}
\end{eqnarray} 
where the infinitesimal quantities $d {\hat \theta}_i$ are due to the ${\hat \zeta}_{\vec{k}}$ mode only (according to (\ref{phi-noV})), and where - as we have discussed - we have disregarded the subdominant contribution to $\delta g_{00}$ and $\delta g_{0i}$ from  ${\hat \zeta}_{\vec{k}}$. 

In ref.   \cite{Bartolo:2013msa} the anisotropy was computed in a coordinate system in which ${\hat \phi}^i = x^i$. We can do the same by rescaling the spatial coordinates by different factors before and after the time $t_k$. To compute the effect of the change of the anisotropy at $t_k$, it is actually convenient to simply redefine the coordinates in (\ref{after}) in such a way that the scalar field profile does not contain the $d {\hat \theta}_i$ contribution. In these coordinates, the line element in (\ref{after}) changes into 
\begin{eqnarray}
&& d s^2 = - d t^2 + {\rm e}^{2 \alpha - 4 \sigma} \, \left( \frac{ 1 + {\hat \theta}_1 }{ 1 + {\hat \theta}_1 + d {\hat \theta}_1 } \right)^2 d x^2 \nonumber\\
&& \quad\quad\quad +  {\rm e}^{2 \alpha + 2 \sigma} \, \left( \frac{ 1 + {\hat \theta}_2 }{ 1 + {\hat \theta}_2 + d {\hat \theta}_2 } \right)^2 \left[ d y^2 + d z^2 \right] \,, 
\end{eqnarray}
(where for brevity we do not change the name of the coordinates). 

The prefactors that originate from $d {\hat \theta}_i$ can be reabsorbed in a change of $\alpha$ and $\sigma$, so that eqs. (\ref{after}) can be replaced by 
\begin{eqnarray}
&& {\hat \phi}^1 = \left( 1 + {\hat \theta}_1 \right) x \;, \nonumber\\ 
&& {\hat \phi}^2 = \left( 1 + {\hat \theta}_2 \right) y \;,\; 
{\hat \phi}^3 = \left( 1 + {\hat \theta}_2 \right) z \;,\; \nonumber\\ 
&&  ds^2 =  - d t^2 + {\rm e}^{2 \left( \alpha + d {\hat \alpha}_{\vec{k}} \right)  - 4 \left( \sigma + d {\hat \sigma}_{\vec{k}} \right) } d x^2 \nonumber\\
&& \quad\quad\quad\quad +  {\rm e}^{2 \left( \alpha + d {\hat \alpha}_{\vec{k}} \right) + 2 \left( \sigma + d {\hat \sigma}_{\vec{k}} \right) } \left[ d y^2 + d z^2 \right] \,, 
\label{after2}
\end{eqnarray} 
with 
\begin{eqnarray}
d {\hat \alpha}_{\vec{k}} &=& - \frac{d {\hat \theta}_1}{3 \left( 1 + {\hat \theta}_1 \right)} -  \frac{2 \, d {\hat \theta}_2}{3 \left( 1 + {\hat \theta}_2 \right)} + {\rm O } \left( d {\hat \theta}_i^2 \right) \nonumber\\ 
&=& - \frac{d {\hat \theta}_1}{3} -  \frac{2 \, d {\hat \theta}_2}{3} + {\rm O } \left( {\hat \theta}_i ,\,  d {\hat \theta}_i^2 \right) \,, \nonumber\\ 
d {\hat \sigma}_{\vec{k}} &=&   \frac{d {\hat \theta}_1}{3 \left( 1 + {\hat \theta}_1 \right)} -  \frac{ d {\hat \theta}_2}{3 \left( 1 + {\hat \theta}_2 \right)} + {\rm O } \left( d {\hat \theta}_i^2 \right) \nonumber\\ 
&=&  \frac{d {\hat \theta}_1}{3} -  \frac{d {\hat \theta}_2}{3} + {\rm O } \left( {\hat \theta}_i ,\,  d {\hat \theta}_i^2 \right) \,. 
\end{eqnarray} 

The comparison between (\ref{before}) and (\ref{after2}) shows immediately the effect of the mode ${\hat \zeta}_k$ on the anisotropy $\sigma$. We note that also $\alpha$ gets modified by the IR effects (in the same way as a quantum jump of a scalar field also modifies $H$ in stochastic inflation), however this overall change does not affect the anisotropy, and is moreover negligible over the change of $\sigma$ (since isotropy requires~\footnote{In standard models, for a flat universe, the normalization of the scale factor is unphysical. We note that this is no longer the case in solid inflation, once the vevs (\ref{solid-bck}) are imposed at the initial time. This is because eqs. (\ref{solid-bck})  relate physical scalar fields to comoving coordinates. Small anisotropy in solid inflation requires both $\sigma \ll \alpha $ and $\dot{\sigma} \ll \dot{\alpha}$. } $\sigma \ll \alpha$).

The result $d {\hat \sigma}_{\vec{k}} =   \frac{d {\hat \theta}_1 - d {\hat \theta}_2 }{3} $ is the initial condition for the anisotropy induced by the mode ${\hat \zeta}_k$. This initial condition is set when the mode 
 ${\hat \zeta}_k$ has become classical, namely at some moment right after horizon crossing. The precise moment at which we set it is irrelevant, since ${\hat \zeta}_{\vec{k}}$ is nearly frozen outside the horizon. While ${\hat \zeta}_{\vec{k}}$ outside the horizon is exactly constant in the standard case, in solid inflation it evolves in a slow-roll suppressed manner, $\frac{d {\hat \zeta}_{\vec{k}}}{d t } = {\rm O } \left( \epsilon H \right) \times  {\hat \zeta}_{\vec{k}}$  \cite{Endlich:2012pz}. Therefore, without the need to explicitly compute the precise coefficient, we write 
\begin{eqnarray}
d {\hat \sigma}_{\vec{k}} \Big\vert_{t_k} &=& \frac{ d {\hat \theta} }{3} \;, \nonumber\\ 
\frac{d }{d t} d {\hat \sigma}_{\vec{k}} \Big\vert_{t_k} &=& \gamma \, \epsilon\, H \,  d {\hat \theta} \;, 
\label{sigmak-in}
\end{eqnarray}
where $ d {\hat \theta} \equiv  d {\hat \theta}_1 -  d {\hat \theta}_2$, and where $\gamma$ is an order one coefficient. After $t_k$ the anisotropy $d {\hat \sigma}_{\vec{k}} $ due to the mode ${\hat \zeta}_{\vec{k}}$ evolves classically according to (\ref{sigma-class}). The two conditions (\ref{sigmak-in}) allow us to determine the integration constants $\sigma_1$ and $\sigma_2$ in  (\ref{sigma-class}). We are only interested in $\sigma_2$, as this is the coefficient of the mode with slowly decreasing anisotropy. Disregarding the mode $\propto \sigma_1$, and the slow roll correction to $\sigma_2$, we can finally write 
\begin{equation}
d {\hat \sigma}_{\vec{k}} \left( t \right) = \left\{ 
\begin{array}{l}
0 \quad\quad\quad\quad  \quad\quad\quad\quad  \quad\quad 
,\;\; t < t_k \;\;, \\
\frac{ d {\hat \theta} \left( t_k \right) }{3}   \, {\rm e}^{- \int_{t_k}^t \left( 1 + c_L^2 \right) \,  \epsilon H d t'} \;\;,\;\; t > t_k \,. 
\end{array} \right. 
\end{equation} 
or
\begin{equation}
d {\hat \sigma}_{\vec{k}} \left( t \right) = \frac{ d {\hat \theta} \left( t_k \right) }{3}  \, \left[ \frac{ a \left( t_k \right)}{a \left( t \right)} \right]^{\epsilon \left( 1 + c_L^2 \right)} \, \vartheta \left( t - t_k \right) \;, 
\label{ds-k}
\end{equation}
where $\vartheta$ is the Heaviside step function. 

As we mentioned, at the linearized level the anisotropies from every IR mode  ${\hat \zeta}_{\vec{k}}$ add up to each other. Therefore, from (\ref{phi-noV}) and (\ref{ds-k}) we obtain the anisotropy at the time the CMB modes leave the horizon
\begin{equation}
{\hat \sigma } \left( t_{\rm CMB} \right) \!\! = \!\! \int_{k_{\rm in}}^{k_{\rm CMB}} \!\!\!\!\!\! \frac{d^3 k}{\left( 2 \pi \right)^{3/2} } \!\! 
\left( {\hat k}_1^2 - {\hat k}_2^2 \right) \! {\hat \zeta}_{\vec{k}} \left( t_k \right) \! \left[ \frac{a \left( t_k \right)}{a \left( t_{\rm CMB} \right)} \right]^{\epsilon \left( 1 + c_L^2 \right)} \,. 
\end{equation} 

The variance of this quantity is given by 
\begin{eqnarray}
&& \left\langle {\hat \sigma} \left( t_{\rm CMB} \right)^2 \right\rangle  =  \frac{4}{15} \, \int_{k_{\rm in}}^{k_{\rm CMB}}  \frac{d k}{k} P_\zeta \left( k \right) \, \left[ \frac{ a \left( t_k \right) }{ a \left( t_{\rm CMB} \right) } \right]^{2 \epsilon \left( 1 + c_L^2 \right)} \nonumber\\ 
&& \quad\quad\quad = \frac{4}{15} \, P_\zeta \left( k_{\rm CMB} \right) \, \frac{1- {\rm e}^{- \left[ n_s -1+2 \epsilon \left( 1 + c_L^2 \right) \right] \left( N_{\rm tot} - N_{\rm CMB} \right)}}{n_s-1+2\epsilon\left(1+c_L^2\right)} \;. \nonumber\\  
\end{eqnarray} 

Using the central value Planck result  \cite{Ade:2013uln} $n_s - 1 =-0.04$ and $P_\zeta \left( k_{\rm CMB} \right) \simeq P_\zeta \left( k_* \right) =  2.2 \cdot 10^{-9}$, we finally obtain 
\begin{eqnarray}
\sigma_{\rm expected} &=& \sqrt{ \left\langle {\hat \sigma} \left( t_{\rm CMB} \right)^2 \right\rangle  }
\nonumber\\ 
& \simeq & 2.4 \cdot 10^{-5} \, \sqrt{\frac{1-{\rm e}^{-\left[2 \epsilon \left( 1 + c_L^2 \right) - 0.04 
\right] \left( N_{\rm tot} - N_{\rm CMB} \right)}}{2 \epsilon \left( 1 + c_L^2 \right) - 0.04 }} \;, \nonumber\\ 
\label{sigma-exp}
\end{eqnarray}
that is the most natural value for the asymmetry at the time in which the CMB modes were produced.

\section{Statistical anisotropy in the CMB}
\label{CMB}
The background anisotropy $\sigma$ gives rise to a violation of statistical anisotropy of the primordial perturbations. Under the simplifying assumption of SO(2) residual symmetry done in the previous section 
(see the paragraph before eq. (\ref{before})), the power spectrum of the primordial perturbations is of the ACW form \cite{Ackerman:2007nb} 
\begin{equation}
P_\zeta \left( \vec{k} \right) = P_\zeta \left( k \right) \left[ 1 + g_* \, \cos^2 \theta_{{\hat k},{\hat i}} \right] 
\,, 
\label{acw}
\end{equation}
with \cite{Bartolo:2013msa,Endlich:2013jia} 
\begin{equation}
g_* = \frac{20}{3} \, \frac{F_Y}{F} \, \frac{1}{c_L^2 \, \epsilon} \sigma = 3 \, c_2 \, \sigma \,, 
\label{gs-c2}
\end{equation} 
(notice that the quantity $\sigma$ introduced in \cite{Endlich:2013jia} corresponds to our $-2 \sigma$) where $c_2$ is the NG coefficient given in eq. (\ref{NG-c2}). As we wrote in that equation, $c_2 \la 60$ at $95 \% \, {\rm C.L.}$ from the Planck data. From eq. (\ref{sigma-exp}) we then obtain 
\begin{eqnarray}
\frac{g_{*,{\rm expected}}}{c_2/60} &\simeq& 4.4 \cdot 10^{-3} \,  \nonumber\\
&& \quad \times  \, \sqrt{\frac{1-{\rm e}^{-\left[2 \epsilon \left( 1 + c_L^2 \right) - 0.04 
\right] \left( N_{\rm tot} - N_{\rm CMB} \right)}}{2 \epsilon \left( 1 + c_L^2 \right) - 0.04 }} \;, \nonumber\\ 
\label{gs-exp}
\end{eqnarray} 
where we have normalized $c_2$ to its maximal possible value.

\begin{figure}
\centerline{
\includegraphics[width=0.5\textwidth]{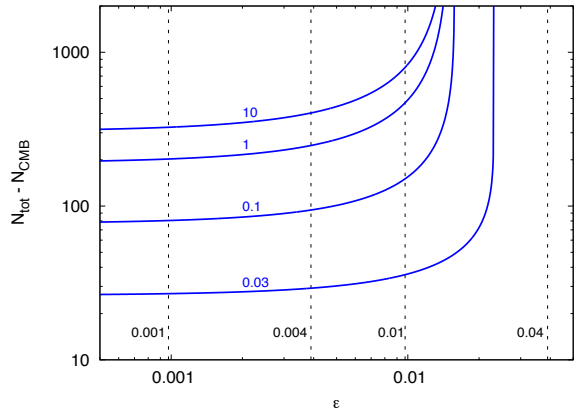}
}
\caption{Blue solid lines: ratio (\ref{gs-exp}). The expected anisotropy in solid inflation is given by this quantity times $\frac{c_2}{60}$ (where $c_2$ parametrizes the amount of non-gaussianity in the model, and 
 $\frac{c_2}{60} \leq 1$ at $95\%$ C.L.). Black dashed lines: tensor-to-scalar ratio $r$. The $x-$ axis is  the standard slow roll parameter $\epsilon$, while the $y$ axis is the duration of inflation in addition to the final $N_{\rm CMB} \simeq 50-60 $ e-folds. The value $c_L = \frac{1}{\sqrt{3}}$ is assumed in this figure. 
}
\label{fig:fig1}
\end{figure}

In Figure \ref{fig:fig1} we show the ratio (\ref{gs-exp}) in solid blue lines, as a function of $\epsilon$ and $N_{\rm tot} - N_{\rm CMB}$. We fix $c_L = \frac{1}{\sqrt{3}}$ (the maximal possible value) in the Figure. The result (\ref{gs-exp}) depends smoothly on $c_L$ only through the combination $1+c_L^2$, and we verified that no qualitative change in  (\ref{gs-exp}) takes place as $c_L$ is decreased towards $0$.  We recall that  the expected anisotropy is obtained by multiplying the value of the ratio (\ref{gs-exp}) shown in the figure by $\frac{c_2}{60}$. As we discussed, this factor is phenomenologically constrained to be $\leq 1$. 

We observe from the figure that the ratio (\ref{gs-exp}) grows with  $N_{\rm tot} - N_{\rm CMB}$ (at any fix value of $\epsilon$) and decreases with $\epsilon$ (at any fix value of $N_{\rm tot} - N_{\rm CMB}$). This 
property can also be verified analytically from (\ref{gs-exp}). The growth with $N_{\rm tot} - N_{\rm CMB}$ 
is immediate to understand, as it reflects the fact that, with a longer duration of inflation, more IR modes are produced, leading to a greater anisotropy. The dependence on $\epsilon$ is instead due to the classical behavior of the anisotropy due to each individual IR mode. As seen in eq. (\ref{ds-k}), this contribution decreases with time with a rate $\propto \epsilon \left( 1 + c_L^2 \right)$. This decrease is more effective the smaller the momentum of the mode is, as, the smaller momentum is, the longer is the time during which this mode is outside the horizon. This effect is compensated by the red tilt of the spectrum, so that the smaller $k$ is, the greater is the initial value of  $d {\hat \theta} \left( t_k \right)$ at horizon crossing. This explains the $2 \epsilon \left( 1 + c_L^2 \right) - \left( n_s - 1 \right)$ dependence in eq. (\ref{gs-exp}). When this combination is positive (hence, when $\epsilon$ is sufficiently large), the decrease with time that mostly affects the very small $k$ modes dominates over the fact that they start with a larger value. In this case, increasing $N_{\rm tot} - N_{\rm CMB}$ leads only to a very small increase of 
 (\ref{gs-exp}), as the most IR modes do not contribute significantly to the anisotropy. The opposite takes place when  $2 \epsilon \left( 1 + c_L^2 \right) - \left( n_s - 1 \right) < 0$, in which case increasing the number of IR modes results in an (exponentially) strong increase of the anisotropy.

\begin{figure}
\centerline{
\includegraphics[width=0.5\textwidth]{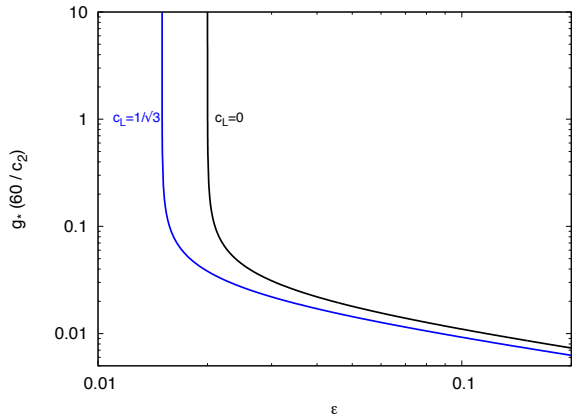}
}
\caption{Asymptotic value of the ratio (\ref{gs-exp}) at asymptotically large $N_{\rm tot} - N_{\rm CMB} \rightarrow \infty$, as a function of the slow roll parameter $\epsilon$, and for the two limiting values of $c_L$.  
This asymptotic value diverges at $\epsilon < \frac{1-n_s}{2 \left( 1 + c_L^2 \right)}$. 
}
\label{fig:fig2}
\end{figure}

This behavior can also be seen in Figure \ref{fig:fig2}, where we show the  value of the ratio (\ref{gs-exp}) at asymptotically large $N_{\rm tot} - N_{\rm CMB} \rightarrow \infty$. We show this quantity as a function of $\epsilon$ and for the two limiting values $c_L = 0$ and $c_L = \frac{1}{\sqrt{3}}$. We see from (\ref{gs-exp}) that this asymptotic value if finite only for  $2 \epsilon \left( 1 + c_L^2 \right) - \left( n_s - 1 \right) > 0$. 

As we explained in the previous paragraph, in this regime modes with progressively smaller $k$ contribute less and less to the expected asymmetry, so that the ratio  (\ref{gs-exp}) converges to a finite value even when an infinite amount of IR modes ($k_{\rm in} \rightarrow 0$) is included in the sum. In the opposite regime (this is the region at the left of the asymptotic vertical lines) modes with progressively smaller $k$ contribute more and more to the anisotropy, and the  ratio  (\ref{gs-exp}) grows exponentially with  $N_{\rm tot} - N_{\rm CMB}$. 

In  Figure \ref{fig:fig1}  we also show in black dashed lines the value of the tensor-to-scalar ratio $r$ in the model, which is given by 
\begin{equation}
r = 16 \, \epsilon \, c_L^5 \,. 
\label{r}
\end{equation} 
Contrary to the mild dependence on $c_L$ of the ratio (\ref{gs-exp}), the value of $r$ instead strongly decrease if smaller values of $c_L$ are considered.

\section{Conclusion}
\label{conclusion}
Elastic / Solid inflation \cite{Gruzinov:2004ty,Endlich:2012pz} is an effective description of inflation that  differs from the vast majority of inflationary models for its peculiar symmetry breaking: individual sources in solid inflation break isotropy, but they are combined to give overall isotropic expansion. This peculiar properties give rise to specific signatures that are not found in standard inflationary models: an anisotropic squeezed bispectrum  \cite{Endlich:2012pz} and the possibility of a prolonged anisotropic background solution \cite{Bartolo:2013msa}. Both features were previously obtained in the $f \left( \phi \right) F^2$ mechanism, in which a suitable function of the inflaton allows for a constant vev of some vector field(s) \cite{Watanabe:2009ct,Barnaby:2012tk}. 
 
Both in solid inflation and in the $f \left( \phi \right) F^2$ mechanism the perturbation of the individual anisotropic sources are (nearly) frozen and scale invariant outside the horizon. As pointed out in 
\cite{Bartolo:2012sd} this unavoidably leads to violation of statistical isotropy. In the  $f \left( \phi \right) F^2$ mechanism, one can consider a theory ${\cal L} \supset - \frac{f \left( \phi \right)}{4} F_{\mu\nu}^{(a)}  F^{\mu\nu,(a)} $ with the global SO(3) symmetry $A^{(a)} \rightarrow R^a_b \, A^{(b)}$. One can then obtain isotropic expansion if the vev of the three vectors are orthonormal
\begin{equation}
\langle A_x^{(1)} \rangle = \langle A_y^{(2)} \rangle = \langle A_z^{(3)} \rangle \,. 
\label{vector-bck}
\end{equation}

 The SO(3) invariance of the starting action ensures that the classical equations of motion preserve this orthonormality. However, the quantum fluctuations of the three fields are completely independent of each other and they need to respect the SO(3) symmetry only on average. If we could observe a large number ${\cal N}$ of realization of inflation, and compute the distributions of the three vector fields in all these realizations, then the three  distributions would be identical in the ${\cal N} \rightarrow \infty$ limit. However, we can observe only one realization, and the fluctuations do not obey any SO(3) symmetry. Vector field perturbation $\delta \vec{A}$ that leave the horizon before the CMB modes add up to a constant value on our Hubble patch. For all phenomenological purposes, we cannot distinguish them from a coherent value, and - even if we assume that the vector field vevs are given by  (\ref{vector-bck}) at some given moment during inflation, they modify the vector vevs observed in our Hubble patch  from   (\ref{vector-bck}) to a non-isotropic value. We cannot deterministically obtain this value from the theory; however we can evaluate the expected value of their stochastic addition. This quantity increases with the amount of inflation before the CMB modes left the horizon. 

Ref.  \cite{Bartolo:2012sd} performed this computation for the  $f \left( \phi \right) F^2$ mechanism, showing that even a few ($\sim 10$) number of e-folds of inflation results in a too large breaking of statistical isotropy \cite{Kim:2013gka}. 
However it was mentioned there that some level of anisotropy necessary takes place in any model in which individual anisotropic sources have a (nearly) scale invariant spectrum. As much as the fluctuations of the individual vector fields break (\ref{vector-bck}) in any one given realization of the $f \left( \phi \right) F^2$ mechanism, the fluctuations of the individual scalar fields of solid inflation break the isotropic vev configuration (\ref{solid-bck}). In this work, we have quantified the expected level of statistical anisotropy due to this breaking in the solid inflation model. 

%
%
%

Statistical anisotropy is parametrized by the ACW \cite{Ackerman:2007nb} parameter $g_*$, that multiplies the anisotropic part of the power spectrum, see eq.  (\ref{acw}). In solid inflation this quantity is proportional to the background anisotropy $\sigma$ times the amplitude of the squeezed bispectrum. The squeezed bispectrum is a measure of how a large wavelength mode can influence the power spectrum at shorter scales. The dependence of $g_*$ on it is due to the fact that  the anisotropy of the power spectrum can be understood \cite{Endlich:2013jia} as the modulation of the two point function of the CMB perturbations from the much larger IR modes that produce the background anisotropy $\sigma$. Therefore it is possible that, even if the IR sum provides a large background anisotropy, the effect in the power spectrum is small if the parameters of the model result is a small non-gaussianity. On the other hand, for the model not be simply an interesting mathematical construction one should hope that we will be able to detect the peculiar shape to its bispectrum. Therefore, if the non-gaussianity in the model is closed to its maximal allowed region, the solid curves  in Figure \ref{fig:fig1} provide a good  estimate on the expected statistical anisotropy in the model. We see that a nontrivial region of the parameter space of the model predicts a level of anisotropy above the observational limit \cite{Kim:2013gka}. Therefore, an interesting breaking of violation of statistical isotropy together with a nonstandard squeezed bispectrum, is a possible signature for this model.

\vskip.25cm
\noindent{\bf Acknowledgements:} 
We thank Sabino Matarrese for useful discussions and for collaboration in the related works \cite{Bartolo:2012sd,Bartolo:2013msa}. We  thank Alberto Nicolis for useful discussions. The work of M. P. is partially supported from the DOE grant DE-SC0011842  at the University of Minnesota. The work of C.U. is supported by a summer grant from the graduate program of the School of Astronomy and Physics of the University of Minnesota. The work of NB was partially
supported by the ASI/INAF Agreement 2014-024-R.0 for the Planck LFI Activity of Phase E2.

\end{document}